\documentclass[showpacs,twocolumn,prl,aps,color]{revtex4}
\usepackage{graphicx}

\begin{document}

\title{Why a local moment induces an antiferromagnetic ordering: An RVB
picture}
\author{Tao Li$^1$, Qiang-Hua Wang$^2$, Zheng-Yu Weng$^3$, Fan Yang$^4$, and
Fu-Chun Zhang$^5$}
\affiliation{$^1$Department of Physics, Renmin University, Beijing, 100874, China\\
$^{2}$National Laboratory of Solid State Microstructures, Nanjing
University, Nanjing 210093, China\\
$^3$Center for Advanced Study, Tsinghua University, Beijing, 100084, China\\
$^4$Department of Physics, Beijing Institute of Technology, Beijing, 100874,
China\\
$^5$Centre of Theoretical and Computational Physics, and Department of
Physics, University of Hong Kong, Hong Kong, China}
\date{{\small \today}}

\begin{abstract}
Based on a Gutzwiller projected BCS wavefunction, it is shown that a local $%
S=1/2$ moment is present around a vacancy site (zinc impurity) in a form of
staggered magnetic moments, which is a direct consequence of the
short-ranged resonating-valence-bond (RVB) pairing in the spin background.
\end{abstract}

\maketitle

Experimentally, it has been observed that a zinc impurity in the high-$T_{c}$
cuprates generally induces a \emph{free }magnetic moment in its
neighborhood, which contributes to an additional Curie-like $1/T$ signature
in NMR and NQR data \cite{Julien,Alloul,Alloul2,NQR1,NQR2}. In particular,
the distribution of such a free moment around the impurity is in the form of
staggered (AF) magnetic moments at the copper sites. Namely, if the free
moment is polarized, say, by an external magnetic field, its spatial
distribution surrounding the zinc atom looks like an antiferromagnetic
ordering of spins, with the total sum equal to the magnitude of the free
moment.

The induction of a local moment by a nonmagnetic zinc impurity illustrates
the strongly correlated nature of the pure system and poses a great
challenge to various high-$T_{c}$ theories. Here two issues are involved.
One is why a free moment should be spontaneously induced by a nonmagnetic
zinc impurity. The second is that given the fact that a free moment is
trapped nearby, why its distribution exhibits a staggered profile around the
zinc impurity.

There have been a number of articles on the microscopic origin of the
zinc-induced local moment in the framework of the doped Mott insulator
approaches\cite{ZQWang,Liang,Qi}. Wang and Lee~\cite{ZQWang} and Liang and
Lee~\cite{Liang} attribute the induced local moment to a precursor effect of
the AF ordering instability in the pure system. They consider the
antiferromagnetic or spin density wave fluctuation in an RVB state nearby
the zinc impurity. In this picture, the local moment is described by a local
spin density wave order parameter which provides an explanation of the
staggered profile of the local moment \cite{ZQWang,Liang}. Qi and Weng~\cite%
{Qi} used a bosonic RVB theory to study a sudden approximation for the zinc
impurity state, in which an electron is annihilated from the zinc-free RVB
state. In their picture, the staggered distribution of the local moment
induced by the zinc impurity is a result of the short range AF correlations
embedded in the pure system. They also argued that the phase string effect
may protect the local moment for a topological reason.

In this paper, we model the Zn-impurity by a Gutzwiller projected d-wave BCS
wave function~\cite{anderson,az} in the sudden approximation, in which the
zinc doped system is constructed simply by locally removing an electron at
the zinc site from the pure system. We show that a free moment trapped
nearby a Zn- impurity will generally exhibit a staggered AF distribution,
whereas the unprojected wave function does not. We show an explicit relation
between the magnetic moment in the Zn-impurity state and the spin-spin
correlation in the impurity-free RVB state. We have also examined a more
sophisticated Zn-impurity RVB state, in which the BCS mean field state
before the Gutzwiller projection is modified with some relaxation of the
surrounding RVB background in the presence of an impurity (vacancy) site.
The result is in further support of the sudden approximation. Therefore, the
local staggered moment in the Zn-impurity state may be viewed as a
consequence of the staggered spatial spin-spin correlation functions in the
Zn-impurity free state. From this point of view, the staggered magnetic
moments observed in experiments may not be necessarily related to the AF
instability. Since the Gutzwiller projection is essential to the staggered
spin-spin correlation function, the Zn-impurity effect is a direct probe of
the intrinsic RVB state in the cuprate superconductors.

We start by approximately representing the pure system by a Gutzwiller
projected d-wave BCS state
\begin{equation}
\left\vert \Psi _{0}\right\rangle =P_{G}|\mathrm{BCS}\rangle  \label{pbcs}
\end{equation}%
where $P_{G}=\prod\nolimits_{i}(1-n_{i\uparrow }n_{i\downarrow })$ enforces
the no double occupancy constraint with $n_{i\sigma }$ denoting the electron
number at site $i$ in a two-dimensional square lattice. The Gutzwiller
projection transforms the singlet Cooper pairs in $|\mathrm{BCS}\rangle $
into neutral spin RVB pairs at half-filling which persist into finite hole
doping. Here the mean field state $|\mathrm{BCS}\rangle $ is derived from a
renormalized mean field Hamiltonian~\cite{Baskaran,Zhang88,Kotliar} such as
the Hamiltonian given in Eq. (4) below.

Now we construct a new state $\left\vert \Psi _{0}\right\rangle _{\mathrm{Zn}%
}$ by removing \emph{locally} an electron on site $i_{0}$ from $\left\vert
\Psi _{0}\right\rangle $. This state may be regarded as a \textquotedblleft
sudden approximation\textquotedblright\ for the \emph{true} ground state $%
\left\vert \Psi \right\rangle _{\mathrm{Zn}}$ when site $i_{0}$ is occupied
by a zinc impurity, similar to a construction previously used in the bosonic
RVB approach\cite{Qi}. We note that $\left\vert \Psi _{0}\right\rangle _{%
\mathrm{Zn}}$ has the right spin quantum number ($S=1/2$) to describe a
local moment. At the same time, both $\left\vert \Psi \right\rangle _{%
\mathrm{Zn}}$ and $\left\vert \Psi _{0}\right\rangle _{\mathrm{Zn}}$ should
reduce to $\left\vert \Psi _{0}\right\rangle $ in a distance sufficiently
far away from the zinc site, as a single impurity is not expected to change
the thermodynamic bulk properties of the system. Thus, we expect $\left\vert
\Psi \right\rangle _{\mathrm{Zn}}$ and $\left\vert \Psi _{0}\right\rangle _{%
\mathrm{Zn}}$ should have a finite overlap such that the local states of
them around the zinc impurity can be adiabatically connected by some kind of
local perturbations, provided the local moment is present in $\left\vert
\Psi \right\rangle _{\mathrm{Zn}}$. Namely, if the $S=1/2$ moment is indeed
distributed around the zinc impurity in the true ground state, then the
\textquotedblleft sudden approximation\textquotedblright\ state $\left\vert
\Psi _{0}\right\rangle _{\mathrm{Zn}}$ should be a good approximation as the
first step to describe the related physics.

Without the loss of generality one may define
\begin{eqnarray}
\left\vert \Psi _{0}\right\rangle _{\mathrm{Zn}}\equiv c_{i_{0}\uparrow
}\left\vert \Psi _{0}\right\rangle .  \label{zn}
\end{eqnarray}%
Since $\left\vert \Psi _{0}\right\rangle $ is a spin singlet state, one has $%
S^{z}=-1/2$ in $\left\vert \Psi _{0}\right\rangle _{\mathrm{Zn}}$, which
forms a doublet with the state $c_{i_{0}\downarrow }\left\vert \Psi _{0}
\right\rangle $. Now let us inspect the distribution of this moment on the
lattice by considering the quantity $\langle S_{j}^{z} \rangle_{Zn}$, the
average of $S_{j}^{z}$ in the state $\left\vert \Psi _{0}\right\rangle _{%
\mathrm{Zn}}$. Let $\langle Q \rangle$ be the average of operator $Q$ in the
pure RVB state $\left\vert \Psi _{0}\right\rangle$, we have for the
translational and rotational invariant RVB state,
\begin{eqnarray}
\langle S_j^z \rangle_{Zn} = \frac{\langle c_{i_0\uparrow}^{\dagger} S_j^z
c_{i_0 \uparrow} \rangle} {\langle c_{i_0 \uparrow}^{\dagger} c_{i_0
\uparrow}\rangle} =\frac{2}{3n}\langle \vec S_{j} \cdot \vec
S_{i_{0}}\rangle,  \label{sm}
\end{eqnarray}%
where $n$ is the average number of electron per site. Therefore, the moment
distribution is determined by the equal-time spin-spin correlation in the
pure system. The right-hand side of Eq.(\ref{sm}) can be numerically
determined by a variational Monte Carlo calculation based on the Gutzwiller
projected BCS state. In a system with antiferromagnetic spin-spin coupling
on a square lattice, the sign of the spin-spin correlation is alternating in
space. Therefore we expect a staggered magnetic moment in the Zn-impurity
state.

To make our discussion more quantitatively, we consider a renormalized
Hamiltonian, from which the mean field ground state of $|\mathrm{BCS}\rangle
$ in Eq. (1) is derived,
\begin{eqnarray}
\mathrm{\mathbf{H}}_{\mathrm{MF}} &=&-t^{\prime}\sum_{\langle ij\rangle
,\sigma }\left( c_{i,\sigma }^{\dagger }c_{j,\sigma }+\mathrm{H.C.}\right)
-\mu \sum_{i\sigma }c_{i,\sigma }^{\dagger }c_{i,\sigma }  \nonumber \\
&&+\sum_{\langle ij\rangle }\Delta _{ij}\left( c_{i,\uparrow }^{\dagger
}c_{j,\downarrow }^{\dagger }-c_{i,\downarrow }^{\dagger }c_{j,\uparrow
}^{\dagger }+\mathrm{H.C.}\right)
\end{eqnarray}%
in which $\Delta _{ij}$ denotes the d-wave pairing order parameter. This
Hamiltonian can be taken as the mean field description of the t-J model,
where the parameters $t^{\prime}$ and $\Delta _{ij}$ can be determined by a
set of self-consistent equations\cite{Baskaran,Zhang88,Kotliar}. The results
are shown in Fig. 1, which illustrates two things. First, there does exist
staggered magnetic moments around the zinc impurity, which are actually
determined by the short-range AF correlations already presented in the \emph{%
pure} system as shown by the right-hand side of Eq.(\ref{sm}); Second, the
total moment as the sum of these staggered moments is equal to $S^{z}=-1/2,$
according to the above construction. Namely the staggered moments spatially
distributed around a zinc impurity and a \textquotedblleft
free\textquotedblright\ $S=1/2$ total moment can be naturally reconciled
within this scheme.

\begin{figure}[h]
\includegraphics[width=5.5cm,angle=0]{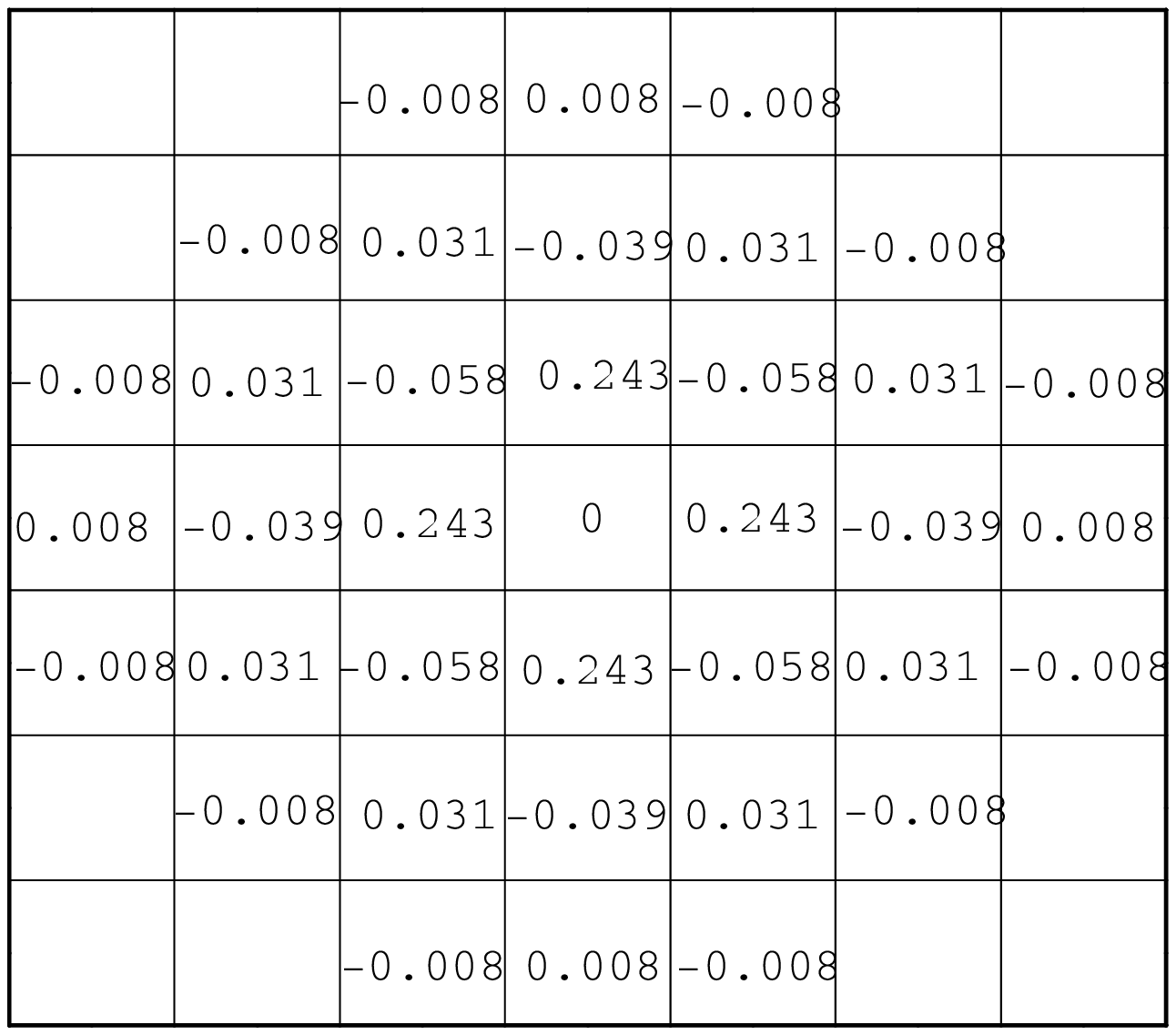}
\caption{The spin distribution of $\left\langle S_{j}^{z}\right\rangle $
around the zinc site (denoted by the plaquette in the center) in the state $%
\left\vert \Psi _{0}\right\rangle _{\mathrm{Zn}}$. The calculation is done
on a $12\times 12$ lattice of 22 holes, with $\frac{\Delta }{t^{\prime }}%
=0.1 $ and $\frac{\protect\mu }{t^{\prime }}= -0.2$.}
\label{fig1}
\end{figure}

Physically, the local moment around the zinc impurity can be understood as
originated from a broken RVB pair, with its one spinon partner (sitting at
the site $i_{0})$ being removed as the consequence of the Zn substitution
and the unpaired partner left behind to contribute to such a free local
moment. Here the Mott physics also plays a crucial role for making such a
free moment distributing in the form of enhanced staggered moments as shown
in Fig. 1. By contrast, without the Gutzwiller projection operator $P_{G}$,
one easily finds that the distribution of magnetic moments is given by

\begin{eqnarray}
&&\frac{\langle \mathrm{BCS}|c_{i_{0}\uparrow }^{\dagger }\left(
S_{j}^{z}\right) c_{i_{0}\uparrow }|\mathrm{BCS}\rangle }{\langle \mathrm{BCS%
}|c_{i_{0}\uparrow }^{\dagger }c_{i_{0}\uparrow }|\mathrm{BCS}\rangle }
\nonumber \\
&=&-\frac{1}{n}\left[ |\langle \mathrm{BCS}|c_{i_{0}\uparrow }^{\dagger
}c_{j\downarrow }^{\dagger }|\mathrm{BCS}\rangle |^{2}+|\langle \mathrm{BCS}%
|c_{i_{0}\uparrow }^{\dagger }c_{j\uparrow }|\mathrm{BCS}\rangle |^{2}\right]
~  \nonumber \\
&\leq &0~  \label{1}
\end{eqnarray}%
which always remains negative and does not show a strong staggered
oscillation like in Fig. 1. So the presence of staggered AF moments around
the zinc impurity is a very important feature for an RVB superconductor
(Gutzwiller projected BCS state) $\left\vert \Psi _{0}\right\rangle $, but
not for a conventional BCS state $|\mathrm{BCS}\rangle $.

The above discussion is based on the assumption that $\left\vert \Psi
_{0}\right\rangle _{\mathrm{Zn}}$ has a finite overlap with the true ground
state $\left\vert \Psi \right\rangle _{\mathrm{Zn}}$. Namely, the existence
of a local moment is assumed to be around a zinc impurity. If the local
moment in $\left\vert \Psi _{0}\right\rangle _{\mathrm{Zn}}$, created by the
\textquotedblleft sudden approximation\textquotedblright\ from the original
ground state $\left\vert \Psi _{0}\right\rangle $, eventually escapes to
infinity (boundary) in the ground state, then $\left\vert \Psi
_{0}\right\rangle _{\mathrm{Zn}}$ and $\left\vert \Psi \right\rangle _{%
\mathrm{Zn}}$ are orthogonal and the present \textquotedblleft sudden
approximation\textquotedblright\ used above will fail. In principle, an RVB
state does not prevent such a local moment (spinon) from escaping away to
the boundary as it can re-arrange the RVB pairings in the bulk. So a more
sophisticated approach is needed in order to further demonstrate the
stability of the local moment around the Zn impurity.

For this purpose, in the following we consider a modified wave function for
describing the local moment. Instead of removing an electron in the
projected state of the \emph{pure} system, the impurity state is constructed
as a Gutzwiller projected state in which the unprojected mean field state is
obtained in the presence of the zinc impurity with a local quasiparticle
excitation. Generally, an s-wave unitary scatter doped into a d-wave
superconductor will induce a low-energy d-wave resonance state around it,
ensured by the d-wave pairing symmetry\cite{Li}, and if we denote the
creation operator of this resonant state as $\gamma _{\sigma }^{\dagger }$,
then the modified wave function for describing the local moment state is
given by $\mathrm{P}_{G}\gamma _{\sigma }^{\dagger }\left\vert \mathrm{BCS}%
\right\rangle _{\mathrm{Zn}}$, in which $\left\vert \mathrm{BCS}%
\right\rangle _{\mathrm{Zn}}$ denotes the mean field ground state in the
presence of the zinc impurity. Note that here the occupied d-wave resonance
state will have the right spin quantum number to describe a local moment and
at the same time any double occupancy of the same local state, which would
destroy the moment, is prohibited by the Gutzwiller projection. The
remaining task is to check if the proposed wave function still exhibits a
staggered distribution of the moment. The mean field description of the
single zinc impurity at site $i_0$ is given by $H_{MF}$ of Eq. (4) with the
impurity site $i_0$ excluded. 

\begin{figure}[h]
\includegraphics[width=5.5cm,angle=0]{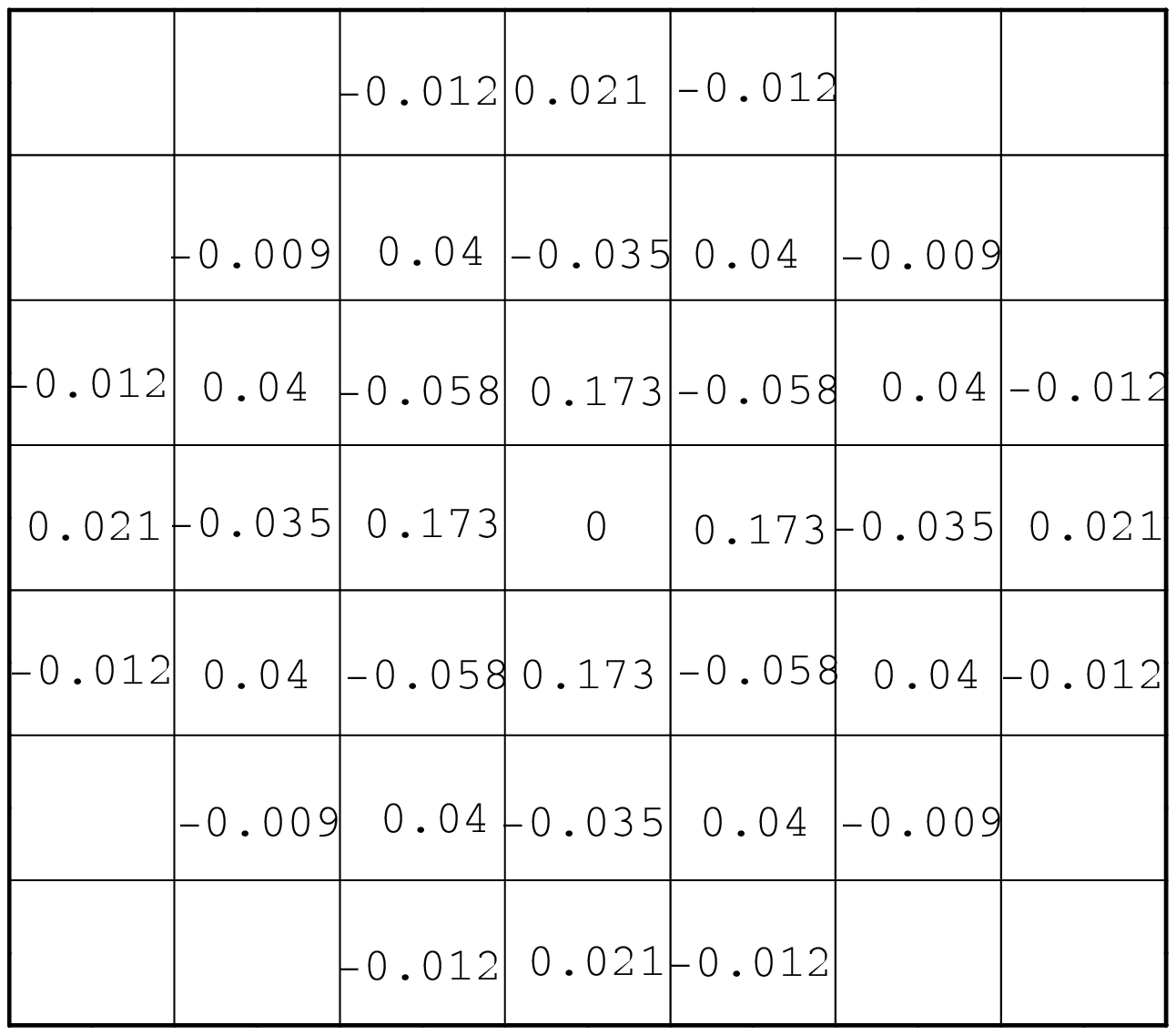}
\caption{The distribution of $\left\langle S_{j}^{z}\right\rangle $ around
the zinc site in the state $\mathrm{P}_{\mathrm{G}}\protect\gamma _{\uparrow
}^{\dagger }\left\vert \mathrm{BCS}\right\rangle _{\mathrm{Zn}}$. The
parameters used are the same as in Fig. 1.}
\label{fig2}
\end{figure}
The mean field ground state of $\mathrm{\mathbf{H}}_{\mathrm{MF}}^{Zn}$ and
the d-wave resonant state $\gamma _{\sigma }$ can be obtained numerically on
the finite lattice. Then one can perform the VMC calculation to implement
the Gutzwiller projection which is done on a $12\times 12$ lattice with 22
holes. Fig. 2 shows the spin distribution around the zinc site in $\mathrm{P}%
_{\mathrm{G}}\gamma _{\uparrow }^{\dagger }\left\vert \mathrm{BCS}%
\right\rangle _{\mathrm{Zn}}$. The spin distribution in $\mathrm{P}_{\mathrm{%
G}}\gamma _{\uparrow }^{\dagger }\left\vert \mathrm{BCS}\right\rangle _{%
\mathrm{Zn}}$ is slightly more extended than that in $\left\vert \Psi
_{0}\right\rangle _{\mathrm{Zn}}$ as a result of the relaxation of the RVB
background but otherwise remains quite similar. Since both $\mathrm{P}_{%
\mathrm{G}}\gamma _{\uparrow }^{\dagger }\left\vert \mathrm{BCS}%
\right\rangle _{\mathrm{Zn}}$ and $\left\vert \Psi _{0}\right\rangle _{%
\mathrm{Zn}}$ have the same spin quantum number and show similar spin
distribution, it is natural to expect that they are not very different from
each other. In fact, we have calculated their overlap by using the VMC and
the result is shown in Fig. 3. From the figure one sees that the overlap
actually saturates to a finite value about 0.8 in the thermodynamic limit.
Thus the two states are indeed very close to each other in terms of both the
large wave function overlap as well as the staggered profile of the local
moment.

\begin{figure}[h]
\includegraphics[width=5.5cm,angle=0]{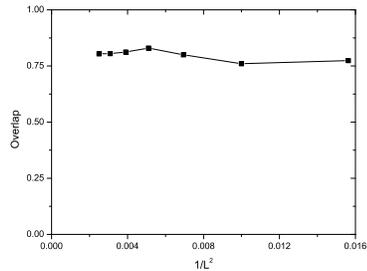}
\caption{The overlap between $\mathrm{P}_{\mathrm{G}}\protect\gamma %
_{\uparrow }^{\dagger }\left\vert \mathrm{BCS}\right\rangle _{\mathrm{Zn}}$
and $\left\vert \Psi _{0}\right\rangle _{\mathrm{Zn}}$ as a function of the
inverse lattice size $1/L^2$. The calculation is done with the hole density
kept approximately at 15 percent (for example, 62 holes in the $20\times 20$
lattice). The parameters, $\frac{\Delta }{t^{\prime }}$ and $\frac{\protect%
\mu }{t^{\prime }},$ are the same as in Fig. 1.}
\label{fig3}
\end{figure}

Energetically a close competitor of the state $\mathrm{P}_{\mathrm{G}}\gamma
_{\uparrow }^{\dagger }\left\vert \mathrm{BCS}\right\rangle _{\mathrm{Zn}}$
is the projected mean field ground state $\mathrm{P}_{\mathrm{G}}\left\vert
\mathrm{BCS}\right\rangle _{\mathrm{Zn}}$. Since the mean field excitation
energy of the resonant state created by $\gamma _{\uparrow }^{\dagger }$ is
small, we expect that the energy difference between $\mathrm{P}_{\mathrm{G}%
}\gamma _{\uparrow }^{\dagger }\left\vert \mathrm{BCS}\right\rangle _{%
\mathrm{Zn}}$ and $\mathrm{P}_{\mathrm{G}}\left\vert \mathrm{BCS}%
\right\rangle _{\mathrm{Zn}}$ be small also. Indeed the energy difference in
the VMC calculation based on the $t-J$ model is generally very small,
although the energy of $\mathrm{P}_{\mathrm{G}}\left\vert \mathrm{BCS}%
\right\rangle _{\mathrm{Zn}}$ remains to be lower. So strictly speaking, how
the local moment can be trapped near the zinc impurity, due to the
distortion of the RVB order parameter around the impurity is still unsolved
in the present approach. In contrast, a topological reason coming from the
charge degrees of freedom in the bosonic RVB theory \cite{Qi} provides the
protection of the local moment by responding nonlocally to the presence of a
vacancy in the bulk. Whether the underlying physics for the trapping of the
local moment is due to the local RVB distortion or some more subtle
topological reason needs to be further studied in the present fermionic RVB
framework. Nevertheless, once a local moment is trapped around the zinc
impurity, a staggered moment profile always emerges, as a direct consequence
of the presence of RVB correlations in the pure system.

In summary, we studied two fermionic RVB wavefunctions describing the zinc
doping-induced local moment effect in the doped Mott insulator. We find that
the AF staggered profile of the local moment distribution emerges naturally
and can be generally attributed to the short-range AF correlations of the
RVB structure in the pure system. According to this description, the
existence of the induced local moment with a staggered profile in the
cuprates provides a strong experimental indication for the underlying RVB
structure in the zinc-free systems.

\begin{acknowledgments}
ZYW thanks X. L. Qi for useful discussions. TL is partially supported by
NSFC Grant No.90303009 and QHW is by NSFC Grant No. 10325416 and the Fok
Ying Tung Education Foundation No.91009, ZYW acknowledges the support from
the grants of NSFC and MOE, and FCZ is partially supported by the RGC grant
in Hong Kong.
\end{acknowledgments}


\begin{thebibliography}{99}
\bibitem{Julien} M.-H. Julien, T. Feh$\acute{e}$r, M. Horvati$\acute{c}$, C.
Berthier, O.N. Bakharev, P. S$\acute{e}$gransan, G. Collin and J.-F.
Marucco, Phys. Rev. Lett 84, 3422 (2000)

\bibitem{Alloul} A. V. Mahajan, H. Alloul, G. Collin, and J.-F. Marucco,
Phys. Rev. Lett. 72, 3100(1994); Eur. Phys. J. B 13, 457 (2000)

\bibitem{Alloul2} W. A. MacFarlane, J. Bobroff, H. Alloul, P. Mendels, N.
Blanchard, G. Collin, and J.-F. Marucco, Phys. Rev. Lett 85, 1108 (2000)

\bibitem{NQR1} G. V. M. Williams and S. Kr$\ddot{a}$mer, Phys. Rev. B 64,
104506 (2001)

\bibitem{NQR2} Y. Itoh, T. Machi, C. Kasai, S. Adachi, N. Watanabe, N.
Koshizuka, and M. Murakami, Phys. Rev. B 67, 064516 (2003)

\bibitem{Chem} R. S. Howland and T. H. Geballe, S. S. Laderman, A.
Fischer--Colbrie, and M. Scott, J. M. Tarascon and P. Barboux, Phys. Rev. B
39, 9017 (1989)

\bibitem{ZQWang} Z. Q. Wang, P. A. Lee, Phys. Rev. Lett 89, 217002 (2002).

\bibitem{Liang} Shi-Dong Liang and T. K. Lee Phys. Rev. B 65, 214529 (2002).

\bibitem{Qi} X.L. Qi and Z. Y. Weng, Phys. Rev. B 71, 184507 (2005).

\bibitem{anderson} P. W. Anderson, Science 235, 1196 (1987)

\bibitem{az} P. W. Anderson, P. A. Lee, M. Randeria, T. M. Rice, N. Trivedi,
and F. C. Zhang, J. Phys.: Condens. Matter \textbf{16}, R755 (2004).

\bibitem{Baskaran} G. Baskaran, Z. Zou, and P. W. Anderson, Solid State
Commun. 63, 973 (1987).

\bibitem{Kotliar} G. Kotliar and J. Liu, Phys. Rev. B 38, 5142 (1988).

\bibitem{Zhang88} F. C. Zhang, C. Gros, T. M. Rice, and H. Shiba, Supercond.
Sci. Technol. 1, 36 (1988).

\bibitem{Li} T. Li, cond-mat/0310439.
\end{thebibliography}
\end{document}